\documentclass[useAMS,usenatbib]{mn2e}
\usepackage{graphicx}
\usepackage{amsmath}
\usepackage{amssymb}
\usepackage{color}

\newcommand{\Mpch}{\mbox{ $h^{-1}$ Mpc}}
\newcommand{\kpch}{\mbox{ $h^{-1}$ kpc}}

\newcommand{\be}{\begin{equation}}
\newcommand{\ee}{\end{equation}}

\def\ltsima{$\; \buildrel < \over \sim \;$}
\def\simlt{\lower.5ex\hbox{\ltsima}}
\def\gtsima{$\; \buildrel > \over \sim \;$}
\def\simgt{\lower.5ex\hbox{\gtsima}}

\title[]{The formation of CDM haloes I: Collapse thresholds and the ellipsoidal collapse model}
\author[Ludlow, Borzyszkowski \& Porciani]{Aaron D. Ludlow\thanks{E-mail: aludlow@astro.uni-bonn.de}, Mikolaj Borzyszkowski \& Cristiano Porciani\\
Argelander-Institut f\"ur Astronomie, Auf dem H\"ugel 71, D-53121 Bonn, Germany}

\begin{document}

\maketitle 
\begin{abstract}
  In the excursion set approach to structure formation initially spherical regions of the
  linear density field collapse to form haloes of mass $M$ at redshift $z_{\rm id}$ if their
  linearly extrapolated density contrast, averaged on that scale, exceeds some critical 
  threshold, $\delta_{\rm c}(z_{\rm id})$. The value of $\delta_{\rm c}(z_{\rm id})$ is 
  often calculated from the spherical or ellipsoidal collapse
  model, which provide well-defined predictions given auxiliary properties of the tidal field
  at a given location. We use two cosmological simulations of structure growth in a $\Lambda$
  cold dark matter scenario to quantify $\delta_{\rm c}(z_{\rm id})$, its dependence
  on the surrounding tidal field, as well as on the shapes of the Lagrangian regions that
  collapse to form haloes at $z_{\rm id}$. Our results indicate that the ellipsoidal
  collapse model provides an accurate description of the mean dependence of
  $\delta_{\rm c}(z_{\rm id})$ on both the strength of the tidal field and on halo mass.
  However, for a given $z_{\rm id}$, $\delta_{\rm c}(z_{\rm id})$ depends strongly
  on the halo's characteristic formation redshift: the earlier a halo forms, the higher its initial
  density contrast. Surprisingly, the majority of haloes forming {\em today} fall below
  the ellipsoidal collapse barrier, contradicting the model predictions.
  We trace the origin of this effect to the non-spherical shapes of Lagrangian haloes,
  which arise naturally due to the asymmetry of the linear tidal field. We show that a modified
  collapse model, that accounts for the triaxial shape of protohaloes, provides a
  more accurate description of the measured minimum overdensities of recently collapsed objects.
\end{abstract}


\section{Introduction}
In the standard model for the growth of large-scale structure, cold dark matter (CDM)
haloes form hierarchically from low-amplitude Gaussian density fluctuations 
seeded during inflation. As the Universe expands, sufficiently over-dense 
regions eventually break away from the Hubble flow, recollapse and form virialized 
objects. Due to their highly non-linear nature, these objects are best studied using
direct $N$-body simulation, and dedicated numerical work has revolutionized our 
understanding of the most fundamental aspects of their formation, evolution and 
structure. Never the less, simulations are often difficult to interpret and insight 
into the rudimentary principles of structure formation can be significantly 
enhanced by supplementing them with simple physical models for the collapse and 
virialization of dark matter (DM) haloes.

One successful analytic approach to modelling structure formation, known as the 
Extended Press-Schechter formalism \citep[e.g.,][hereafter EPS]{PS1974,Bond1991}, relates 
the (Eulerian) statistics of DM haloes to the properties of the Lagrangian 
density field from which they emerge. Although \cite{PS1974} originally used this 
model to calculate the mass function of collapsed objects at different redshifts, 
it has since been extended and used to compute a variety of hierarchical clustering 
statistics, such as halo merger histories 
\citep[e.g.,][]{KauffmannWhite1993,Lacey1993,ShethLemson1999a,vandenBosch2002}, 
and spatial correlations \citep{MoWhite1996,Mo1997,Catelan1998,Porciani1998,ShethLemson1999b}.

The accuracy of the model is surprising given that it overlooks a variety 
of non-linear and scale-dependent aspects of gravitational clustering. Regarding 
structure formation, its simplicity is also its strength: it provides an intuitive 
picture of a much more complex process, and can help to guide and interpret more 
sophisticated numerical calculations of structure growth.

One of the principal assumptions of EPS theory (also known as the excursion-set formalism) 
is the idea that a mass element at position ${\bmath{x}}$ at some high initial redshift
$z_i\gg 0$ will be associated with a halo of mass $M$ at some later redshift, $z_{\rm c}$, 
provided its linear overdensity, $\delta_M({\bmath{x}},z_{\rm c})$ (smoothed on scale $M$ 
and extrapolated to that redshift by linear theory), exceeds a threshold $B$. For an 
initial overdensity $\delta_M({\bmath{x}},z_i)$ at redshift $z_i$, collapse therefore 
occurs provided
\begin{equation}
\delta_M({\bmath{x}},z_i) \, \frac{D(z_{\rm c})}{D(z_i)} = B(z_{\rm c})
\end{equation}
is satisfied, where $D(z)$ is the growth factor of linear perturbations.

The collapse ``barrier'', $B$, however, is not known a priori, and one must gain insight 
from simple dynamical models for gravitational collapse. The spherical collapse (SC) model 
\citep{Peebles1980}, for example, follows the evolution of a sphere of constant overdensity in 
an otherwise unperturbed but expanding background. This model is patently scale-free 
and (for a given collapse redshift, $z_{\rm c}$) the barrier therefore depends only on the 
local density, $B(z_{\rm c})\equiv \delta_{\rm sc}(z_{\rm c})\approx 1.686$. Although this 
result is derived for an Einstein--de Sitter universe, it is approximately valid for 
a large range of cosmologies. \cite{Eke1996} provide useful analytic corrections to the 
value of $\delta_{\rm sc}$ for cosmologies with $\Omega_{\rm M}+\Omega_{\Lambda}=1$.

\citet{White1996} pointed out that EPS theory, combined with the SC
threshold, fails when applied on a point-by-point basis. Using $N$-body simulations, 
they showed that the largest mass associated with a Lagrangian overdensity $\delta_{\rm sc}$ 
at random particle locations correlates poorly with the mass of the halo in which the 
same particles were found at $z=0$. A significant improvement can be made, however, by 
considering the Lagrangian locations $\bmath{x}_h$ of ``protohaloes'' -- regions
that later collapse to form haloes -- rather than random points. However, even with this 
input from simulations, the SC barrier, $B=\delta_{\rm sc}$, still 
systematically over-predicts the corresponding halo masses \citep{Sheth2001}, with a 
discrepancy that increases towards lower mass.

It is well-known that perturbations in Gaussian random fields are inherently triaxial 
and are therefore subject to tidal forces \citep{Doroshkevich1970,Bardeen1986}. This 
motivated attempts to modify the spherical model to incorporate the influence of 
tides on halo collapse. Building upon the classic work of \cite{LyndenBell1964} and 
\cite{LinMestelShu1965}, a number of authors studied the effect of internal shear on 
the collapse of homogeneous ellipsoids in a uniformly expanding background 
\citep[e.g.,][]{Icke1973,WhiteSilk1979,Peebles1980,WatanabeInagaki1991,Lemson1993}. 
\cite{EisensteinLoeb1995}  and \cite[][hereafter BM96]{Bond1996} generalized this model to follow the 
collapse of an initially spherical perturbation in the presence of external shear. In 
this case, the sphere is distorted into an ellipsoid whose principal axes are parallel 
to those of the external tidal field. This model has since become known as the ellipsoidal 
collapse model (hereafter, the EC model) due to the triaxiality of both the tidal field and 
the collapsing ellipsoid itself. Using this model one can ascertain the impact of tidal 
shear on the properties of collapsed regions.

The tidal stretching and compression of initially spherical perturbations in the EC 
model results in a barrier height that depends on all three eigenvalues of the tidal 
deformation tensor at ${\bmath{x}}$. \cite{Sheth2001} realized 
that the ellipsoidal barrier, combined with the statistical properties of Gaussian 
random fields, implied that fluctuations of lower mass should be subject to larger 
degrees of tidal deformation. They used this to approximate the dependence of $B$ 
on halo mass and showed that, when properly tuned, the EC model
employed within the excursion-set formalism provides a more accurate description of the 
abundance of DM haloes, as well as the statistical properties of their progenitor 
populations \citep{Sheth2002}.

Using cosmological simulations of structure growth, 
\citet[see also Dalal et al. 2008; Elia et al. 2012]{Robertson2009} 
measured the Lagrangian overdensities of regions which later collapse 
to form haloes of mass $M$, and found a $\delta(M)$ relation that resembled the expectations 
of the EC model, suggesting that it captures the most relevant 
aspects of gravitational collapse. None the less, when adopting the ellipsoidal collapse 
barrier and the excursion set ansatz for random points, they failed to provide an accurate 
prediction of the DM halo mass function. Indeed, the success of the ellipsoidal 
model in predicting the 
Eulerian statistics of the DM halo population is not based on a direct 
application of $B_{\rm ec}$ within the EPS theory. Instead, a functional form for $B_{\rm ec}$ 
(whose {\em shape} is motivated by the EC model) is determined and 
calibrated directly against the measurable halo statistic that one wishes to predict. 
\citet{Sheth2001}, for example, rescaled the EC barrier in order to reproduce the mass 
function of DM haloes obtained from numerical simulations. It is therefore important 
to critically investigate the EC model and the excursion set ansatz separately, 
in order to assess whether the SC or EC models themselves provide a reasonable physical picture of 
halo formation. 

The purpose of this paper is to investigate the statistical properties of Lagrangian 
DM haloes (or protohaloes) found in $N$-body simulations of structure growth and to 
compare them with the expectations of the spherical and ellipsoidal collapse models. The 
remainder of the paper is organized as follows. In Section~\ref{sec:numerics} we introduce 
our simulations and briefly describe our main analysis techniques. The basic characteristics 
of protohaloes in our simulations are presented in Section \ref{sec:protohaloes}, focusing 
mainly on their Lagrangian overdensities and surrounding tidal fields. Motivated by those 
results, we present a modified EC model in Section~\ref{sec:boundary}. 
Finally, we provide a brief discussion and summary of our findings in Sections
\ref{sec:discussion} and \ref{sec:summary}.

\section{Numerical Methods}
\label{sec:numerics} 

We present here a brief summary of our numerical simulations. More details can be 
found in \citet{Pillepich2010} and \citet{LudlowPorciani2010}.

\subsection{Simulations}
\label{ssec:simulations} 

Our analysis focuses primarily on DM haloes extracted from two high-resolution
simulations of structure formation in the standard $\Lambda$CDM cosmology. Each run adopted 
the following cosmological parameters: $\Omega_{\rm M}=0.279$, $\Omega_{\Lambda}=1-\Omega_{\rm M}=0.721$, 
$\sigma_8=0.817$, $n_s=0.96$, and $H_0\equiv H(z=0)=73$ km s$^{-1}$ Mpc, consistent with those of 
the {\it Wilkinson Microwave Anisotropy Probe} five-year data release \citep{Komatsu2009}. Here 
$\Omega_i$ is the contribution to the total energy density of the Universe from component $i$; $\sigma_8$ 
is the rms mass fluctuation measured in 8 $h^{-1}$ Mpc spheres, linearly extrapolated to $z=0$; 
$n_s$ is the spectral index of primordial density fluctuations, and $H_0$ is Hubble's constant.

The linear density fields for each run were sampled using $1024^3$ equal-mass particles in 
periodic boxes with side-lengths equal to $l_{\rm box}=150$ and $1200$\Mpch. 
For these choices of cosmological parameters, box sizes, and particle number the particle 
masses are 2.433$\times 10^8$ and 1.246$\times 10^{11}$ $h^{-1}$ M$_{\odot}$ 
in the $150$ and $1200$\Mpch{} boxes, respectively. The corresponding softening lengths 
are 3 and 20\kpch, which are kept fixed in comoving coordinates throughout the simulations.

Initial conditions for our simulations were generated by perturbing the initially uniform 
distribution of particles using the Zel'dovich approximation consistent with a starting 
redshift of $z_{\rm i}=70$ for the 150\Mpch{} box and $z_{\rm i}=50$ for the 1200\Mpch{} box. 
As discussed by \citet{Pillepich2010}, these redshifts are sufficiently high so that possible 
transient features in the halo mass function are effectively erased by $z\approx 2$.
For each simulation, 30 snapshots of the particle distribution were saved between $z=10$ and 0 
in equally spaced steps of $\log\, (1+z)^{-1}$.

\subsection{Halo and Protohalo Catalogues}
\label{ssec:haloes} 

In each run, DM haloes were identified in all 30 simulation outputs using a 
friends-of-friends (FoF) halo finder with a link length of 0.2 times the mean nearest-neighbour
spacing. We retain the identities of all FoF haloes that contain more than 32 particles but, in 
what follows, we restrict our analysis to those having at least ${\rm N_{\rm min}}=500$. We have 
verified that our results are insensitive to our particular choice of halo finder by repeating 
the most pertinent aspects of the analysis using DM haloes identified with a 
spherical-overdensity halo finder that adopted a density contrast of $200$ times the mean 
matter density, $\rho_{\rm m}$.

Our analysis focuses on DM haloes identified at four separate redshifts: $z_{\rm id}=0$, 
1, 2 and 3. For all haloes identified at $z_{\rm id}$ we construct accretion histories by tracing 
their most massive progenitor haloes backwards through all previous simulation outputs. We will 
use $z_{50}$, the redshift at which the most massive progenitor had first assembled 50 per cent of 
its descendants mass at $z_{\rm id}$, as a proxy for each halo's formation redshift. 
Our estimates of $z_{50}$ are obtained by linearly interpolating the mass
accretion histories between simulation snapshots and we have verified that, over the mass and redshift
ranges studied here, they are not unduly influenced by our output sequence. In what 
follows we will refer to this redshift as $z_{50}$, in order to distinguish it from a more general
``collapse'' redshift, $z_{\rm c}$; the corresponding cosmological times will be referred to as 
$t_{50}$, $t_{\rm c}$ and $t_{\rm id}$.

DM protohaloes are defined as the Lagrangian patches in the linear density field that are 
occupied by the subset of particles belonging to each FoF halo at $z_{\rm id}$. The mass of a 
protohalo is therefore equal to the mass of its descendant at $z_{\rm id}$.

\subsection{Analysis}
\label{ssec:analysis} 

We measure protohalo shapes using the following description of their mass distribution 
\citep[e.g.,][]{ColeLacey1996,BailinSteinmetz2005,Bett2007}:
\begin{equation}
  I_{ij}=\sum_{k}\frac{x_{k,i}x_{k,j}}{x_k^2},
  \label{inertia}
\end{equation}
where $x_k$ is the distance to particle $k$ from the protohalo's centre of mass; and 
$i$ and $j$ are the components of $\bmath{x}$ along the Cartesian axes of the simulation.
The factor of $1/x_k^2$ is present in equation~(\ref{inertia}) to prevent particles at 
large distances from dominating the shape estimates, but in practice our results 
are insensitive to this weighting. This matrix can be diagonalized and the principal 
axis lengths, $q_1\geq q_2\geq q_3$, determined from the square roots of its eigenvalues; 
the eigenvectors, $\bmath{i}_i$, define the principal axis frame of the system. Once the 
axis lengths $q_i$ have been found, protohalo shapes can be characterized in terms of 
their ratios: $q_2/q_1$ measures the intermediate-to-major axis ratio, and $q_3/q_1$ 
the minor-to-major axis ratio.

In addition to shapes, we also measure properties of the tidal field in the 
vicinity of each protohalo using the tidal deformation tensor, defined
\begin{equation}
{\mathcal{D}}_{ij}=\frac{\partial^2\Phi}{\partial x_i \partial x_j}.
\label{deform}
\end{equation}
Here $\Phi({\bmath{x}})$ is the peculiar gravitational potential at position 
${\bmath{x}}$, which is related to the density contrast by Poisson's equation: 
$\nabla^2\Phi({\bmath{x}})=\delta({\bmath{x}})$. Note that, in practice, 
all fields are evaluated in Fourier space. By diagonalizing ${\mathcal{D}}$ we obtain its
ordered eigenvalues, $\lambda_1\geq\lambda_2\geq\lambda_3$; these determine the 
geometry of the tidal field at ${\bmath{x}}$, and their sign indicates whether 
the flow is inward ($+$) or outwards ($-$). The eigenvectors of ${\mathcal{D}}$, 
$\bmath{d}_i$, determine the orientation of the flow. At any time, $\bmath{d}_1$ 
points along the direction of maximum compression; $\bmath{d}_3$ aligns with the 
axis of minimum compression (or expansion). Note that 
$\delta({\bmath{x}})=\sum_i \lambda_i({\bmath{x}})$, so the tidal field contains 
all information necessary to construct linear overdensities at ${\bmath{x}}$.

It is common to express the eigenvalues of ${\mathcal{D}}$ in terms of the 
ellipticity, $e$, and prolateness, $p$, of the tidal field. These are defined by
\begin{equation}
e = \frac{\lambda_1-\lambda_3}{2\, \delta}
\label{eq:ell}
\end{equation}
and
\begin{equation}
p = \frac{\lambda_1-2\, \lambda_2+\lambda_3}{2\, \delta}.
\label{eq:prl}
\end{equation}
The combination of $(e,p,\delta)$ or $(\lambda_1,\lambda_2,\lambda_3)$ therefore 
provide an equivalent description of the linear tidal field. 

\section {protohaloes in cosmological simulations}
\label{sec:protohaloes}

\begin{figure*}
\begin{center}
\resizebox{18cm}{!}{\includegraphics{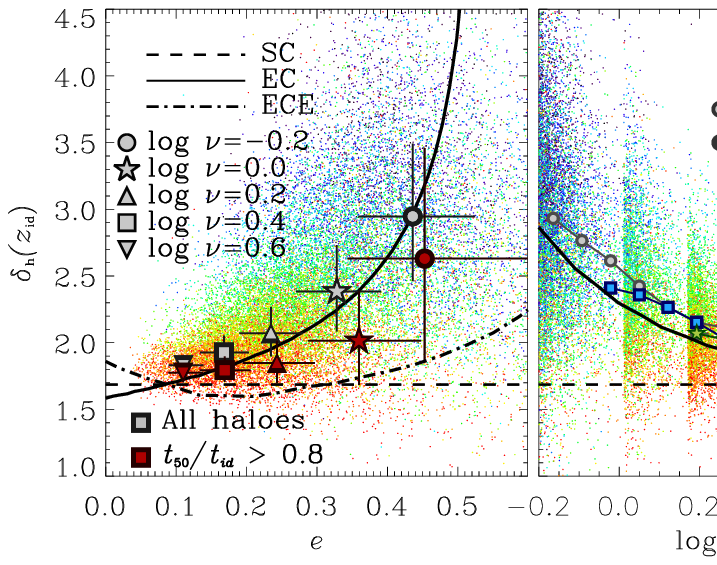}}
\end{center}
\caption{Lagrangian overdensities of protohaloes linearly extrapolated to their identification 
  redshifts, $\delta_{\rm h}(z_{\rm id})$, plotted as a function of the ellipticity of the 
  surrounding tidal field (left) and the dimensionless ``peak height'' mass parameter,
  $\nu(z,M)=\delta_{\rm sc}(z)/\sigma(z,M)$ (right). Light (grey) symbols with error bars in the 
  left-hand panel plot the median $\delta_{\rm h}$ and $e$ for haloes that fall in narrow bins of peak
  height centred on the value indicated in the legend; error bars indicate the 25th and 75th
  percentiles of the scatter. Darker (red) symbols correspond to the medians and scatter for 
  the subsample of haloes in each $\nu$ bin that have $t_{50}/t_{\rm id}>0.8$, corresponding to
  recently collapsed systems. The heavy solid curve shows the predictions of the EC
  model of BM96; the dashed curve corresponds to the collapse threshold for 
  an initially triaxial perturbation with axis ratios consistent with those of $M_\star$ haloes:
  $q_2/q_1=0.75$ and $q_3/q_1=0.55$ (for simplicity, both models assume a prolaticity $p=0$). In 
  the right-hand panel, the same overdensities are plotted as a function of peak height, $\nu$.
  Heavy connected points trace the median relation for haloes identified at different redshifts, 
  $z_{\rm id}$. The good agreement at overlapping $\nu$ indicates that our results are not unduly 
  influenced by numerical artefacts. The thick solid line in the right-hand panel shows the 
  predicted mass dependence of the BM96 EC model assuming that $e=(\sigma/\delta)/\sqrt{5}$
  \citep[e.g.][]{Sheth2001}. In both panels, the horizontal dashed line shows the SC
  barrier, $\delta_{\rm sc}=1.686$.}
\label{Fig:initial_delta_e}
\end{figure*}

\subsection {Overdensities}
\label{ssec:prototides}

The EC model makes specific predictions for the linear overdensity 
required for collapse to occur at redshift $z$. The collapse threshold does not depend 
on any intrinsic property of the halo, but rather on the shape and strength of the 
surrounding tidal field, i.e. on $e$ and $p$. In the left-hand panel of 
Fig.~\ref{Fig:initial_delta_e} we plot the overdensities of protohaloes, $\delta_h$, 
linearly extrapolated to $z_{\rm id}$, as a function of the tidal field ellipticity, $e$. 
Both quantities have been averaged over a spherical Lagrangian volume enclosing the halo 
mass\footnote{We have verified that, for protohaloes with $>$500 particles, 
discreteness effects in our linear density field do not influence our estimates of $\delta_h$ 
and $e$ are negligible (see \citet{Hahn2014}
for an alternative method for measuring these quantities.)}. Individual points correspond 
to haloes (with N$_{\rm FoF}\geq 500$) identified at 
each of the four redshifts mentioned above. The heavy, light-colored symbols plot the 
median values of $\delta_{\rm h}$ and $e$ in several narrow bins of the dimensionless peak 
height parameter\footnote{The choice of peak height, $\nu$, is more natural than mass when 
  comparing haloes across different redshifts. It describes halo masses relative to the 
  characteristic halo mass, $M_\star$, at $z_{\rm id}$, defined by $\nu(M_\star,z_{\rm id})=1$.}, 
$\nu(z,M)=\delta_{\rm sc}(z)/\sigma(z,M)$, as indicated in the legend; the horizontal
and vertical error bars highlight the 25th and 75th percentiles of the scatter along each axis. 
(We have verified that, within the statistical error, the median values of $\delta_{\rm h}$ and $e$ 
at each $\nu$ are independent of $z_{\rm id}$.) 

Note that these points follow very closely the thick solid line, which corresponds to the 
collapse threshold predicted by the EC model of BM96 (assuming $p=0$, the most 
probable value for random points in a Gaussian random field). The agreement between the simulations 
and the EC model is quite remarkable given its simplicity. This supports the model's theoretical 
underpinning, in which more strongly sheared perturbations -- typically those with larger 
ellipticities, $e$ -- require higher initial overdensities to collapse at $z_{\rm id}$.

Never the less, at any $\nu$, the scatter in both $\delta_{\rm h}$ and $e$ is large, and correlates 
strongly with the characteristic formation time of each halo. This can be seen in the colour 
coding of points, which highlight gradients in the formation time variable $\log \, (t_{50}/t_{\rm id})$, 
as indicated in the colour bar on the right. This scatter, and its dependence on $z_{50}$, cannot be 
accounted for by halo-to-halo variation in tidal prolaticity alone \citep[see, e.g.,][]{Hahn2014}
suggesting that other factors are at play.

Note that the solid curve in Fig.~\ref{Fig:initial_delta_e}
was constructed for $p=0$ and assumes $z_{\rm c}=z_{\rm id}$; it therefore represents a lower limit 
to the collapse threshold (for collapse at $z_{\rm c}>z_{\rm id}$, the density threshold would be 
larger by a factor $D(z_{\rm id})/D(z_{\rm c})$). If in reality $z_{\rm c}=z_{\rm id}$, and all 
haloes are collapsing just at the moment they are identified, then their linear overdensities would 
trace the true underlying barrier. Intriguingly, the vast majority of {\em recently collapsed 
haloes lie below the ellipsoidal collapse threshold}. Instead, these haloes sit slightly above the 
spherical threshold, $\delta_{\rm sc}=1.686$, shown as a dashed line in this plot. This is clearly 
at odds with the expectations of the EC model, which predicts that haloes with 
$z_{\rm c}\approx z_{\rm id}$ should trace the solid curve, and those with $z_{\rm c} > z_{\rm id}$ 
to scatter to higher $\delta_{\rm h}$. The heavy (red) filled symbols in the left-hand panel make 
this point clear. These show the median values of $\delta_{\rm h}$ and $e$ (in narrow bins of $\nu$, 
centred on the values indicated in the legend) but for haloes with $t_{50}/t_{\rm id}>0.8$, 
corresponding to recently collapsed systems. Although the tendency of $\delta_{\rm h}$ to increase 
with $e$ remains, the dependence is much weaker than predicted by standard ellipsoidal dynamics: these 
points fall systematically below the solid curve.
More important for EPS theory is the mass dependence of the collapse threshold. We can use the
linear overdensities of our haloes, extrapolated to the halo identification epoch, to relate
$\delta_{\rm h}$ to the halo's final mass at $z_{\rm id}$ (or equivalently, its fluctuation amplitude). 
We show this in the right hand panel of Fig.~\ref{Fig:initial_delta_e}. The thick solid curve shows 
the ellipsoidal collapse threshold predicted by the model of BM96, mapped on to halo mass using the 
most probable $e-\sigma(M)$ relation for random points in a Gaussian random field: 
$e=(\sigma/\delta)/\sqrt{5}$ \citep{Doroshkevich1970,Sheth2001}. 

In agreement with previous work \citep{Dalal2008a,Robertson2009,Elia2012}, we find that the measured 
protohalo overdensities decrease with halo mass in a manner that resembles the predictions of the 
EC model. The agreement, however, is not perfect: at any given mass scale, the 
median Lagrangian overdensity of protohaloes slightly exceeds $\delta_{\rm ec}$. More importantly, 
we find that, for the most recently collapsed systems, the mass dependence of $\delta_{\rm h}$ is 
much weaker than expected from the ellipsoidal model, even though it is precisely these systems that 
we expect to follow the true threshold for collapse at $z_{\rm id}$. 

\begin{figure}
\begin{center}
\resizebox{7.5cm}{!}{\includegraphics{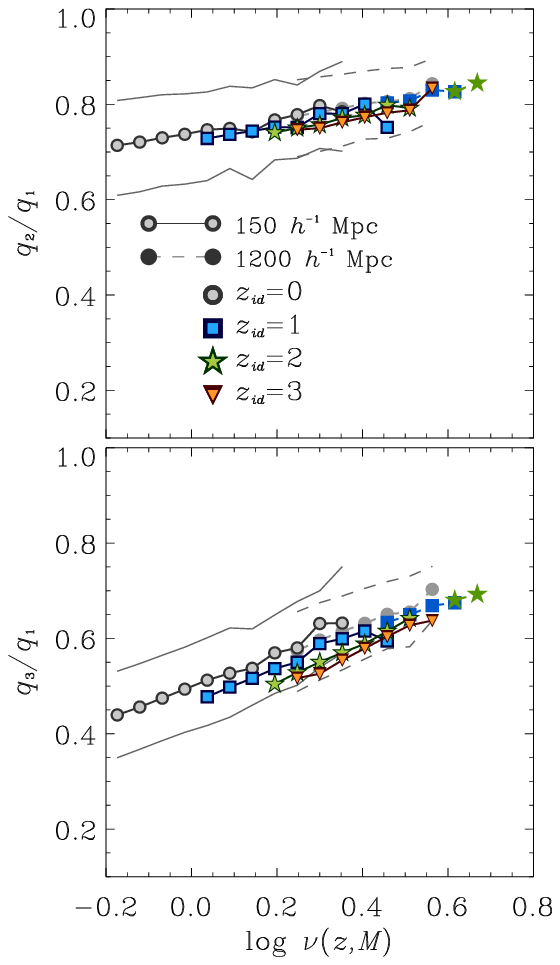}}
\end{center}
\caption{Axis ratios, $q_2/q_1$ (top) and $q_3/q_1$ (bottom), of protohaloes in the initial 
  conditions of our simulations as a function of peak height, $\nu$. Connected points 
  show the median trends, with different symbols used to distinguish protohaloes according 
  to the redshift at which their descendants were identified, $z_{\rm id}$. Outlined points 
  connected by solid lines show results for protohaloes in our 150 $h^{-1}$ Mpc box; solid 
  points for those in our 1200 $h^{-1}$ Mpc box. For $z_{\rm id}=0$, the first and third 
  quartiles of the scatter are indicated using thin solid (150 $h^{-1}$ Mpc box) or dashed 
  (1200 $h^{-1}$ Mpc) lines.}
\label{Fig:initial_shapes_MassDep}
\end{figure}

\subsection {Shapes}
\label{ssec:shapes}

The standard EC model is based on the assumption that protohalo geometries
are perfectly spherical; during collapse, the surrounding tidal field distorts the sphere into 
an ellipsoid and evokes the notion of ``ellipsoidal'' collapse. However, as noted by 
\cite{Porciani2002b}, Lagrangian regions that collapse to form DM haloes by $z=0$ are 
manifestly non-spherical \citep[see also][]{LudlowPorciani2010,Despali2013}. In the EC model, 
initially triaxial perturbations will affect the dynamics of collapse, since the initial 
displacement of the outermost shell (more precisely, the boundary of the perturbation) differs 
along each of the three principal axes of the ellipsoid. This inevitably alters the collapse 
time of each axis and, as a result, the overdensity required for complete collapse to occur by 
a particular time.

\begin{figure*}
\begin{center}
\resizebox{14cm}{!}{\includegraphics{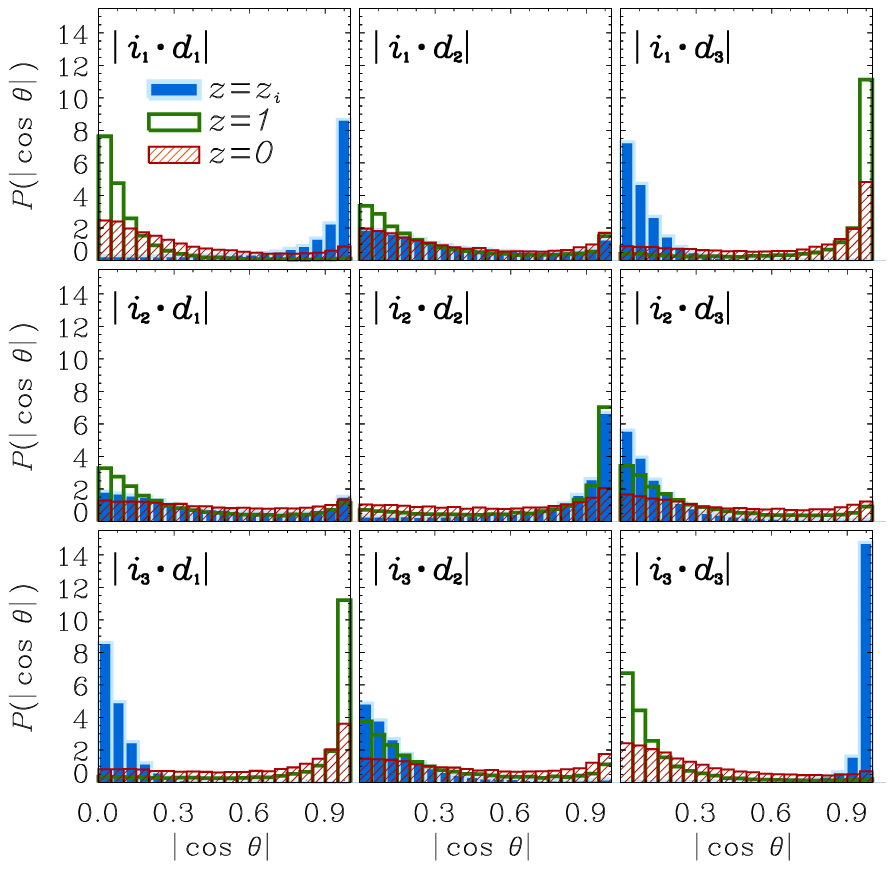}}
\end{center}
\caption{Alignment of the principal axes of the inertia, $I_{ij}$, and deformation, 
  ${\mathcal{D}}_{ij}$ tensors for haloes and protohaloes in our 150 $h^{-1}$ Mpc box. 
  Only haloes in the ($z_{\rm id}=0$) mass range 
  $24.3 < {\rm M_{FoF}}/(10^{10}\,h^{-1} {\rm M}_{\odot}) < 48.7$
  are included. Blue shaded histograms are 
  used for protohaloes in the initial conditions of the simulation; open (green) and 
  (red) hatched histograms show the alignments measured at $z=1$ and $0$, respectively. 
  The eigenvalues are ordered such that $j=1$ corresponds to the major axis and $j=3$ to 
  the minor. The initial alignment between the principal axis frames of the inertia and 
  tidal tensors changes as the system collapses, reflecting changes in the ordering of 
  the eigenvectors of the inertia tensor.}
\label{Fig:alignment}
\end{figure*}

In Fig.~\ref{Fig:initial_shapes_MassDep} we plot the mass dependence (expressed as a function 
of $\nu$) of the initial axis ratios, $q_2/q_1$ and $q_3/q_1$, of protohaloes in both of our 
simulations. Coloured points show separately the median trends for the Lagrangian progenitors
of haloes identified at each $z_{\rm id}$; and full or outlined points
distinguish those identified in our 1200 and 150 $h^{-1}$ Mpc boxes, respectively. 
Thin solid (150 $h^{-1}$ Mpc) and dashed (1200 $h^{-1}$ Mpc) lines in both panels indicate the 
first and third quartiles in each $\nu$ bin (shown only for the $z_{\rm id}=0$ sample,
for clarity). The good agreement between the median trends and scatter at overlapping mass scales 
suggests that our shape estimates are not unduly affected by resolution. 

When expressed in term of peak height, $\nu$, protohalo shapes are largely independent of the
redshift at which their descendants are identified. This suggests that, on average, the rarity of the 
peak, or equivalently its fluctuation amplitude, determines the geometry of the collapsed
region \citep[see also][]{Despali2014}. This is not true for individual haloes, however, and 
we discuss this point further in Section~\ref{ssec:shapeshear}.

More importantly, we find that protohalo shapes depend weakly but systematically on $\nu$, with 
rarer fluctuations becoming increasingly more spherical. We can quantify the departure from 
spherical symmetry by the ``sphericity'', $q_3/q_1$, which varies, on average, from $\sim 0.42$ 
for haloes with $\nu\approx 0.6$ to $\sim 0.65$ for the protohaloes of the rarest, most massive 
haloes at $z_{\rm id}$. Note also that spherical protohaloes are quite rare. Fewer than  
$\sim 0.3$ per cent of haloes with $\nu\approx 0.6$ have $q_3/q_1\geq 0.8$.  This fraction increases 
to $\sim 0.5$ per cent for $\nu\approx 1$, and to $\sim 6.6$ per cent for $\nu\simgt 3$.

\subsection {Orientation}
\label{ssec:orientation}

\begin{figure*}
\begin{center}
\resizebox{16cm}{!}{\includegraphics{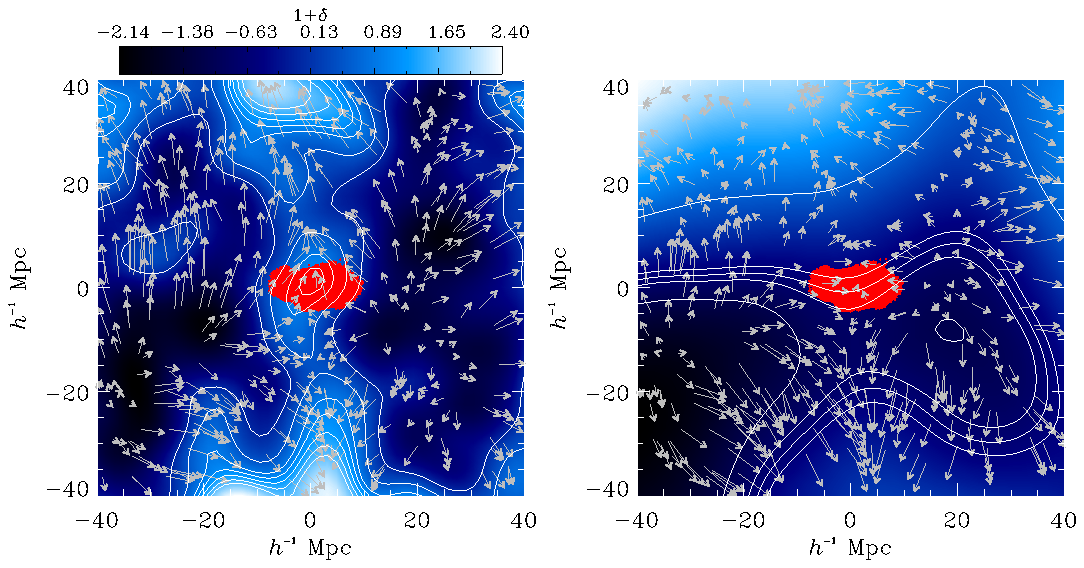}}
\end{center}
\caption{Linear over-density (left) and peculiar gravitational potential 
  (right) fields in the vicinity of a $\sim 3\times 10^{13}\,h^{-1}\,{\rm M}_\odot$ 
  protohalo identified in our 150\Mpch{} box simulation.
  The protohalo is shown using red dots; thin white lines highlight 
  contours of fixed over-density and potential that have been over-laid to 
  guide the eye. Both fields have been smoothed on the halo mass scale and 
  linearly extrapolated to $z=0$. The halo forms in the vicinity of a density 
  peak with the same characteristic mass, but has a markedly different shape
  and orientation. This is because the geometry of the peak is determined by 
  a density threshold, whereas the protohalo shape and orientation are 
  determined by the external tidal field generated by the surrounding large-scale 
  structure. The flow of DM on to and around the protohalo is shown 
  using thin arrows. In the left-hand panel these arrows mark the total bulk velocity 
  field, whereas those on the right show the flow in the halo rest frame.}
\label{Fig:density_field}
\end{figure*}

The orientation of the principal axis frames of protohaloes is closely related to the external tidal
field acting upon them \citep[][see also Lee \& Pen, 2000]{Porciani2002b}. This connection is illustrated
in Fig.~\ref{Fig:alignment}, where we plot the distribution of angles between the three principal
eigenvectors of the inertia, $\bmath{i}_i$, and shear tensors, $\bmath{d}_i$, for protohaloes in 
the ($z_{\rm id}=0$) mass range $24.3 < {\rm M_{FoF}}/(10^{10}\,h^{-1} {\rm M}_{\odot}) < 48.7$.
The different histograms correspond to different times at which the alignment was measured: filled 
(blue) histograms show the alignment measured at $z_i$, the initial redshift of our simulations; the open 
(green) histograms correspond to $z=1$, and the hatched (red) histograms to the final output, at $z=0$. 
Note that, at all times, we track the same set of haloes, and include all particles associated with 
each at $z=0$ in the calculation of the inertia tensor. 

At $z_i$, the peak near $|\cos\theta|\approx1$ in the diagonal panels indicates that the long, 
intermediate and short axes of the principal axis frames of ${\mathcal{D}}$ and $I$ share the 
same orientation \citep[see also][]{Porciani2002b,Despali2013}. The upper-right and lower-left 
panels also show that the minor axis of either tensor lies orthogonal to the major axis of the 
other. The alignment is such that the direction of maximum compression coincides with the major 
axis of the protohalo, whereas its minor axis coincides with the axis of weakest compression, 
or dilation. This coincidence in orientation is actually {\em expected} if the shape of the tidal 
field dictates the Lagrangian region from which particles can accrete by a given time. 

At later times the alignment shifts. At $z=1$, for example, the shortest axis
of the ellipsoid typically aligns with $\bmath{d}_1$, and the longest axis with $\bmath{d}_3$. 
This change in orientation is at odds with the collapse dynamics envisioned by the
BM96 EC model, in which the alignment remains fixed at all times. However, as we will see below, 
the ordering of the principal axes of $I$ may change during collapse if the initial perturbation 
is non-spherical yet perfectly aligned with the principal axes of the deformation tensor: the 
apparent change in orientation reflects a change in the ordering of the 
principal axes frame of the protohalo as it collapses. Two axes switch their
order instantaneously at the moment in which they have equal lengths.
In reality, however, there will always be a small degree of misalignment between
the inertia and shear tensors which will cause a continuous deformation of
the protohalo until ${\bf i}_1$ and ${\bf i}_3$ align with ${\bf d}_3$
and ${\bf d}_1$, respectively \citep{Despali2013}. This initial misalignment
will also generate a small net angular momentum in the collapsing ellipsoid. Finally, 
note that the alignment is still evident at $z=0$, but less so. This is a result of the more 
spherical geometry of virialized systems, and the fact that internal shear (generated by the 
non-spherical halo shape itself) dominates the tidal field due to the halo's high density contrast.

\subsection {Relationship between shape, shear and orientation}
\label{ssec:shapeshear}

If boundaries of protohaloes are determined by the external tidal field then both the 
relative magnitude and sign of the eigenvalues, $\lambda_i$, establish their over-all shape. 
In Fig.~\ref{Fig:density_field} we show an example of this, where we plot the linearly 
extrapolated density field, $1+\delta$, centred on a $\sim 3\times 10^{13}\, h^{-1}\, {\rm M}_{\odot}$
protohalo (red dots) in the initial conditions of our 150\Mpch{} box simulation. The contours 
enclosing the density peak around which the halo collapses provide an intuitive picture of the spatial 
distribution of DM in its vicinity: its mass distribution is extended along a high-density 
ridge connecting two massive structures and separating two low-density voids. The long axis of the 
{\em protohalo}, however, lies transverse to that of the density peak, and it is easy to understand 
why. Material flows out of the voids towards the higher density filament along an axis roughly 
perpendicular to it, whereas the large-scale overdensities result in dilation along the ridge. 
The net effect is a push-and-pull of material that sets the shape and orientation of the protohalo 
such that its long axis coincides with the direction of maximum compression, or infall, and its 
short axis to the direction of maximum expansion. The grey arrows in Fig.~\ref{Fig:density_field} 
sample the linear velocity field and help clarify the impact of large-scale tides on gravitational 
collapse around the density peak.

The fact that the mass distribution associated with the density peak differs substantially
from that of the protohalo indicates that the tidal forces associated with the surrounding
large-scale structure dominate over the internal shear generated by the peak's mass distribution.
In the right-hand panel of Fig.~\ref{Fig:density_field} we show the smoothed potential
field in the same region. Thin white lines show iso-potential contours and arrows the velocity 
field measured in the halo rest frame. Clearly the largest potential gradient lies along the 
shortest axis of the protohalo, with the long axis roughly bounded by iso-potential contours. 
The shapes of protohaloes are thus determined by asymmetry in the large-scale tidal field rather 
than by the shapes of the peaks form which they form. 

\begin{figure}
\begin{center}
\resizebox{8.5cm}{!}{\includegraphics{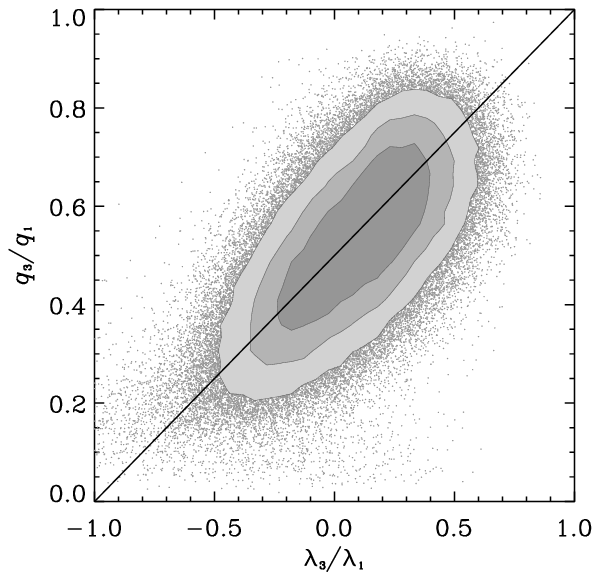}}
\end{center}
\caption{Minor-to-major axis ratios, $q_3/q_1$, of protohaloes versus the ratio
  $\lambda_3/\lambda_1$ of the tidal deformation tensor measured at their centres 
  of mass. Contours enclose 50, 75 and 90 per cent of the data, with the remaining 
  10 per cent shown using grey dots. For clarity, we only plot haloes identified at
  $z_{\rm id}=0$ that also contain more than 500 particles.}
\label{Fig:shape_tides}
\end{figure}

In Fig. \ref{Fig:shape_tides} we plot the ratio $q_3/q_1$ versus $\lambda_3/\lambda_1$ for all 
protohaloes (corresponding to DM haloes identified at $z_{\rm id}=0$) in our simulations (that contain at 
least 500 particles). The filled contours enclose 50, 75 and 90 per cent of the data, with the remaining 
haloes shown individually as dots. The strong correlation between the protohalo sphericity and the 
asymmetry of surrounding tidal field can be understood as follows. The limiting spherical case, in 
which all $\lambda_i$s are equal, corresponds to an isotropic tidal field. In this case there is no 
preferred orientation of the tidal forces acting upon linear perturbations and collapse occurs roughly 
isotropically, resulting in $q_3/q_1\approx q_2/q_1\approx 1$.
A decrease in $\lambda_3/\lambda_1$, however, corresponds to larger dilation along the minor
axis of the tidal field, $\bmath{d}_3$, relative to the enhanced compression along $\bmath{d}_1$. 
This drives material away from the perturbation along $\bmath{d}_3$ while boosting accretion along 
$\bmath{d}_1$. This results in a more flattened geometry (i.e., lower values of $q_3/q_1$) 
as material from larger initial displacements can reach the centre along $\bmath{d}_1$ in a given time.
As we will see in the next section, these correlations have important implications for collapse 
thresholds inferred from the EC model.

\begin{figure}
\begin{center}
\resizebox{8.5cm}{!}{\includegraphics{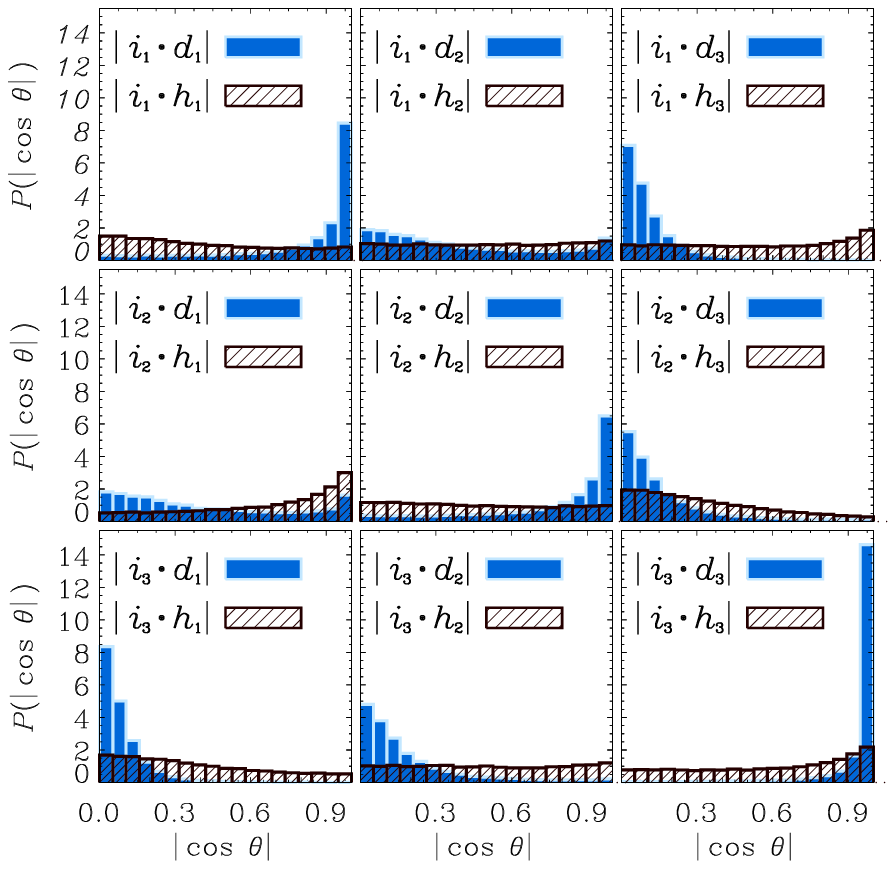}}
\end{center}
\caption{Alignment of principal axes of the protohalo inertia tensor, $I_{ij}$, with 
  those of the tidal field (filled blue histograms) and of the Hessian of the overdensity field
  (black hatched histograms). The latter quantities are measured at protohalo centres of mass
  and have been averaged over a spherical Lagrangian region containing the halo mass. As in
  Fig.~\ref{Fig:alignment}, only haloes in the ($z_{\rm id}=0$) mass range 
  $24.3 < {\rm M_{FoF}}/(10^{10}\,h^{-1} {\rm M}_{\odot}) < 48.7$ are included and, in all 
  cases, the eigenvalues are ordered such that $j=1$ corresponds to the major axis and $j=3$ 
  to the minor axis. As anticipated, the peak shapes are oriented randomly with respect to 
  the protohalo mass.}
\label{Fig:alignment_DelHess}
\end{figure}

These results cast doubt on the validity of previous work in which the Lagrangian volumes 
of protohaloes are assumed to coincide with isodensity contours drawn around local maxima in the 
linear density field. The density distribution in the vicinity of a peak located at $\mathbf{x_p}$ can be
approximated by a Taylor expansion about the peak:
\begin{equation}
  \delta(\mathbf{x_p}) \simeq \delta_0-\frac{1}{2}\frac{\partial^2\delta}{\partial x_i \partial x_j} (x-x_p)_i (x-x_p)_j+...,
\label{eq:delhess}
\end{equation}
where $\partial_i\partial_j\delta$ is the Hessian matrix of the density field. This matrix can be diagonalized;
its eigenvalues and eigenvectors, denoted $\bmath{h}_i$, describe the shape and orientation of the peak, 
respectively. \citet{Heavens1988} used this description of the peak's mass distribution to estimate the total mass 
associated with a given density maxima, while \citet{Catelan1996} used the misalignment between the 
eigenvectors of the peak and those of the surrounding tidal field to calculate the growth of angular 
momentum in proto-galactic haloes.

We explore the validity of this assumption in Fig.~\ref{Fig:alignment_DelHess}, where we plot, 
using black hatched histograms, the distribution of angles between the principal axis frames of the 
protohalo inertia tensor and those of the density Hessian at the protohalo location. Only ($z_{\rm id}=0$) 
haloes in the mass range $24.3 < {\rm M_{FoF}}/(10^{10}\,h^{-1} {\rm M}_{\odot}) < 48.7$ are included. For
comparison, we also show the alignment between the eigenvalues of $I_{ij}$ and of $D_{ij}$ (blue
filled histograms). These results indicate that protohaloes are essentially randomly oriented with
respect to the underlying matter distribution, which strengthens our conclusion that the large-scale
tidal field ultimately dictates the shape and orientation of protohalo regions.

\section {Thresholds from ellipsoidal collapse}
\label{sec:boundary}

In order to better understand these results we here critically examine the EC model, 
bearing in mind the findings presented above. In particular, we describe a general model for 
homogeneous ellipsoidal collapse that follows the evolution of an {\em initially triaxial} overdensity 
in the presence of an evolving external tidal field. We will refer to this model as ECE 
(for Ellipsoidal Collapse of Ellipsoidal perturbations) in order distinguish it from the standard EC 
model. We will use it to study the impact of protohalo shapes on collapse thresholds inferred from 
ellipsoidal collapse dynamics. The dynamical equations for the ellipsoid are derived in full detail 
in Appendix~\ref{sec:ellipsoidal}.

\begin{figure}
  \begin{center}
    \resizebox{8.5cm}{!}{\includegraphics{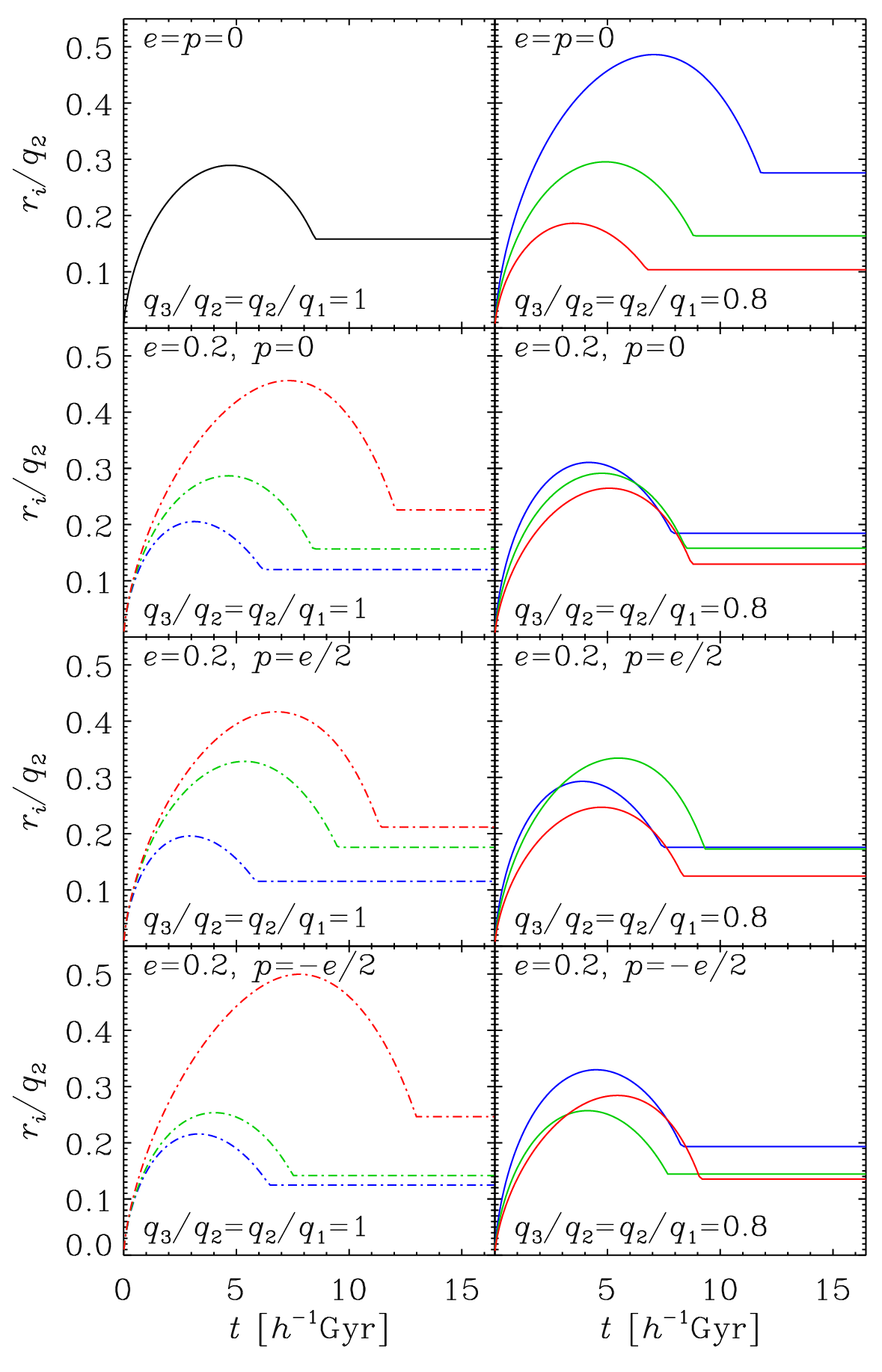}}
  \end{center}
  \caption{Evolution of the axis lengths for collapsing ellipsoids, expressed in units of 
    the initial intermediate axis length, $q_2$. Left-hand panels assume that the
    initial overdensity is a uniform sphere, as in the model of BM96; right-hand panels adopt 
    an initially triaxial overdensity with axis ratios $q_2/q_1=q_3/q_2=0.8$. Top panels assume 
    negligible tidal forces. Lower panels have a (total) tidal field ellipticity of $e=0.2$, 
    and show results for three different prolaticities, $p=0$ (the characteristic value for 
    random points in a Gaussian random field), as well as $p=e/2$ and $p=-e/2$, as indicated 
    in the legends. In each case, the quoted $e$ and $p$ correspond to the {\em total} tidal 
    field. Note that relaxing the assumption of spherical symmetry substantially changes 
    the collapse times of each axis. Axes are color-coded according to their
    alignment with the initial tidal field: blue curves align with $\bmath{d}_1$, green with
    $\bmath{d}_2$ and red with $\bmath{d}_3$.}
  \label{Fig:trajectory}
\end{figure}

\subsection{Homogeneous Ellipsoidal Collapse}
\label{ssec:ellipmodel}

We consider a uniform density perturbation in an otherwise unperturbed Friedmann--Robertson--Walker 
background whose energy content is dominated by the matter density, $\rho_{\rm m}$, and a 
cosmological constant, $\Lambda$. We model the perturbation as a homogeneous ellipsoid with 
semi-axes of physical length $r_i$ ($i=1,2,3$) and density contrast, $\delta$. We adopt a 
Cartesian coordinate system, $x_i$, that aligns with the principal axes of the ellipsoid, as 
well as with the external tidal field (see Fig.~\ref{Fig:alignment}).

In this coordinate system, the axis lengths of the Lagrangian ellipsoid, $r_i$, will obey the 
following equation of motion:
\begin{equation}
\frac{\ddot{r}_i}{r_i}=\frac{\Lambda c^2}{3}-4\,\pi\, G\, \rho_{\rm m} \left(\frac{1+\delta}{3}
+\frac{\beta_i}{2}\,\delta+\lambda^{\rm ext}_i \right)\;.
\label{eq:motion}
\end{equation}
Here $c$ is the speed of light and $\lambda^{\rm ext}_i$ is the $i$th eigenvalue of the 
{\em external} tidal field. The $\beta_i$ characterizes the internal shear generated by 
the shape of the perturbation and can be calculated using standard elliptic integrals 
(e.g., equation~\ref{ellpot} in the Appendix).

\begin{figure}
\begin{center}
\resizebox{8cm}{!}{\includegraphics{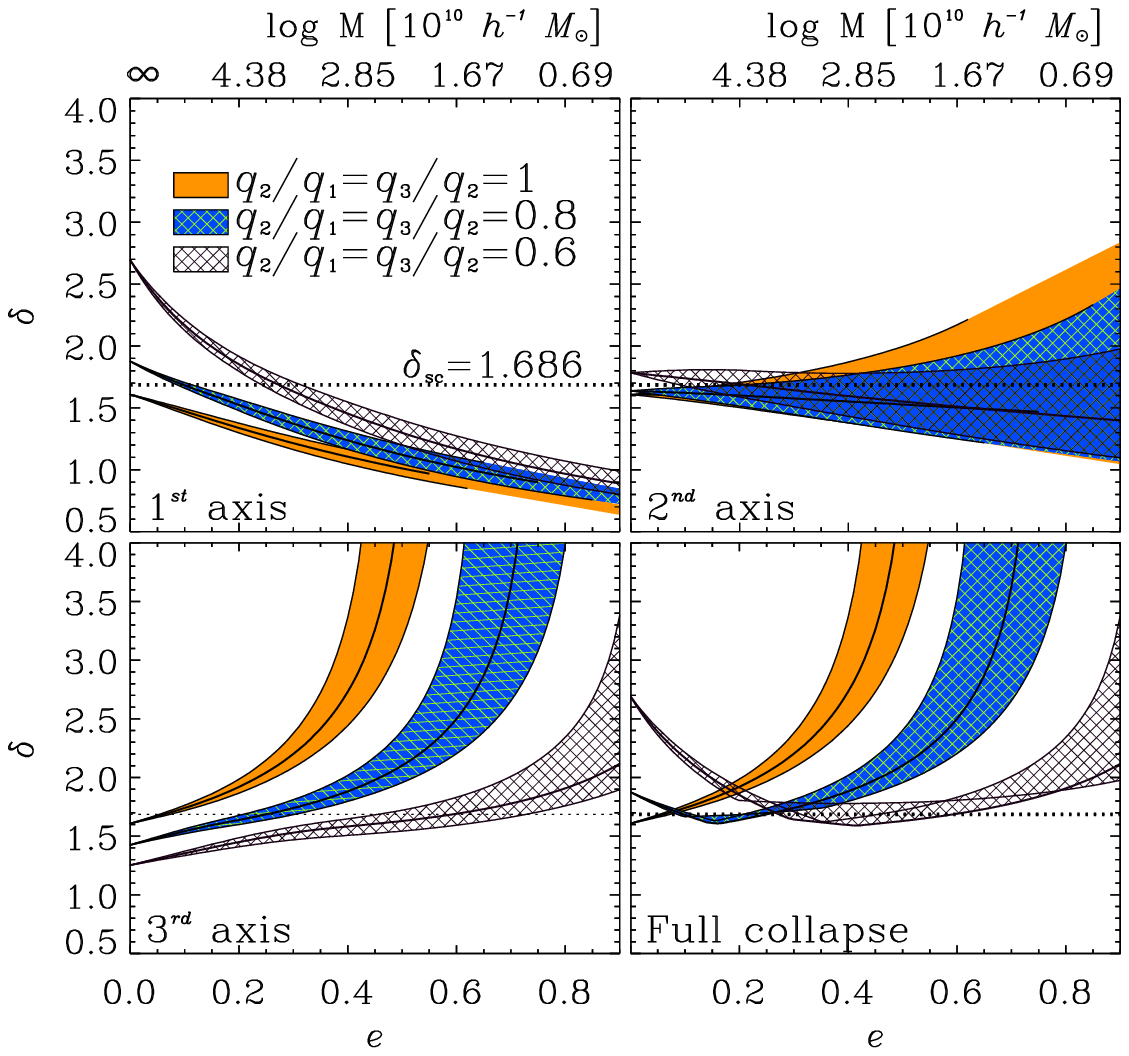}}
\end{center}
\caption{Comparison of the collapse boundaries predicted by the EC
  model for three initial perturbation shapes. The different panels 
  show separately the linearly extrapolated over-density required for collapse 
  to occur along the primary axis (top-left), the intermediate axis (top-right) 
  and the minor axis (bottom-left) of the initial perturbation. The bottom-right 
  panel shows the barrier height required for all three axes to collapse by the 
  present day. The shaded and hatched regions demonstrate the effect of varying 
  the prolaticity of the tidal field over the range $-e/2\leq p \leq e/2$. Orange, 
  blue and white regions show results for three initial shapes: $a_2/a_1=a_3/a_2=1$,
  $a_2/a_1=a_3/a_2=0.8$, and $a_2/a_1=a_3/a_2=0.6$, respectively. In all cases, 
  the major axis of the tidal field is assumed to be aligned with the major axis 
  of the initial perturbation; the corresponding intermediate and minor axes are 
  also aligned.}
\label{Fig:BoundaryCompare}
\end{figure}

Once the initial external tidal field has been specified, solving equation~(\ref{eq:motion})
requires a prescription for its subsequent time evolution. We assume that external tides are 
dominated by large-scale structure and approximate their evolution according to linear theory,
i.e. $\lambda^{\rm ext}_i(t)\propto\lambda_{i,0}^{\rm ext} D(t)$. The internal tides associated 
with the evolving perturbation are calculated self-consistently within the model and depend, at any
moment, on its overdensity and shape.

Because the potential is quadratic and the acceleration is linear in the 
coordinates, concentric ellipsoids with the same axis ratios evolve self-similarly; 
the perturbation therefore remains homogeneous at all times and satisfies
\begin{equation}
1+\delta=\frac{q_1\,q_2\,q_3}{r_1\,r_2\,r_3}\, (1+z)^{-3}\, ,
\label{eq:ovrd}
\end{equation}
where the $q_i$s denote the {\em initial} comoving principal axis lengths of the ellipsoid.
The density contrast grows with time, diverging when any one axis fully collapses. 
In reality, however, small initial departures from perfect symmetry are enhanced during
collapse, and result in a stable structure close to virial equilibrium (BM96). We approximate 
this virialization process by halting collapse along each of the three axes once they reach 
a fraction $f$ of their initial comoving radius. Note that the value of $f$ has no fundamental 
physical meaning, but $f=0.178$ reproduces the virial overdensity of $\sim \,$178 for spherical 
collapse in an Einstein--de Sitter universe. For simplicity, we will also adopt $f=0.178$ in our 
EC model, but note that the final overdensities of haloes in this case will
differ depending on the details of collapse.

In Fig.~\ref{Fig:trajectory} we plot the trajectories of the three axes lengths, in physical 
units, for the EC model of BM96 (left-hand panels) and for the 
modified ECE model described above (right). We start our numerical calculations at $t_0=0.250$ 
Myr and assume, in each case, the same initial density contrast.
The top two panels show the evolution of an initially spherical (left) and 
ellipsoidal perturbation with axis lengths $a_2/a_1=a_3/a_2=0.8$ (right) when no external 
tides are present. The lower panels show, for the same initial set-up, the effect of varying 
the external tidal field on the evolution of the axis lengths.

Notice how relaxing the assumption of spherical symmetry at $t_0$ changes substantially the 
evolution of the axis lengths, even when the tidal field is chosen to be identical. In particular, 
the ordering of the principal axes change with time; what is initially the primary axis, for example, 
can swap between intermediate and minor, and then back to primary again. This explains the 
time-dependence of the alignment between the principal axes of the inertia and shear tensors 
seen in Fig. \ref{Fig:alignment}. Note that this reordering of the principal axes $\bmath{i}_i$ 
is impossible to achieve when linear perturbations are assumed to be spherical; in this case
their orientation at all $t>t_0$ depends entirely on the shape of the initial external tidal field.
 
In addition, note that the turnaround and freeze-out times also differ for initially 
triaxial perturbations.
As shown in Fig.~\ref{Fig:BoundaryCompare}, this has important implications 
for modelling collapse boundaries for halo formation. In each panel, orange, blue and 
white-hatched regions show the collapse thresholds for perturbations with initial axis ratios
$q_2/q_1=q_3/q_2=1$, $0.8$ and $0.6$, respectively. Thick solid lines show results 
for $p=0$; the shading and hatching demonstrates the effect of varying the tidal prolaticity 
over the range $-e/2\leq p \leq e/2$. Each panel plots separately the linearly extrapolated 
density contrast required for collapse to occur today along one of the three principal axes of the 
initial perturbation. (The first axis corresponds to what is initially the major axis of 
inertia; the third to the minor axis.) 

For a given $e$ and $p$, collapse along the major axis of an initially non-spherical perturbation 
always requires a higher linear density contrast than for an initial sphere. This is because the internal 
shear generated by an ellipsoidal perturbation always suppresses collapse along the perturbation's 
major axis, which is, additionally, located at a larger initial displacement than 
in the spherical case. Along the minor axis of the initial ellipsoid, however, collapse thresholds
are always lower than in the spherical case. This is because the internal shear acting along this 
axis always aids in collapse, and large values of tidal ellipticity (corresponding to large and 
negative values of $\lambda_3$) are required to counteract this. 

The lower right panel of Fig.~\ref{Fig:BoundaryCompare} shows the ellipsoidal threshold, $B_{\rm ec}$, 
required for all three perturbation axes to collapse at $z_{\rm c}=0$. Clearly, even modest changes in initial 
shape of the perturbation affects the inferred barrier considerably. For example, the heavy dot--dashed 
line in the left panel of Fig. \ref{Fig:initial_delta_e} shows $B_{\rm ec}$ computed for a perturbation
with axis ratios $q_2/q_1=0.75$ and $q_3/q_1=0.55$, approximately equal to the mean values for haloes
with masses $\sim M_\star$. Now essentially {\em all} haloes lie {\em above the ellipsoidal collapse barrier}, 
which provides an accurate {\em lower bound} the typical overdensities of recently collapsed systems.

\subsection{Shape dependence of barrier heights}
\label{ssec:shapeB}

\begin{figure*}
\begin{center}
\resizebox{18cm}{!}{\includegraphics{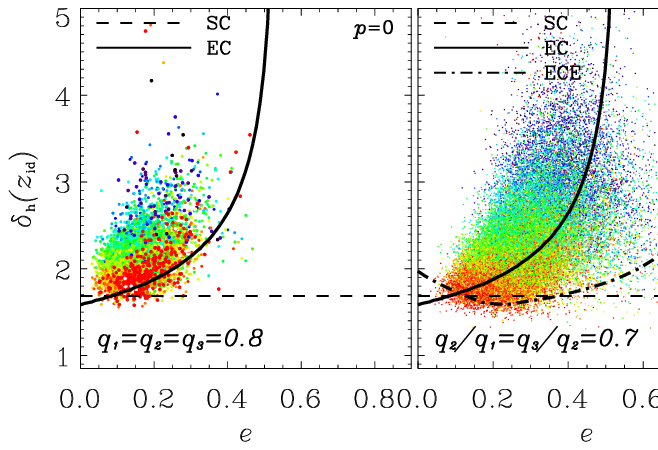}}
\end{center}
\caption{Linear overdensities of protohaloes, $\delta_{\rm h}(z_{\rm id})$, as a function of the
  ellipticity of the surrounding tidal field, $e$, for haloes in three separate shape bins.
  Left-hand panels show ``spherical'' protohaloes, with $q_3/q_1>0.8$; middle and 
  right-hand panels show subsamples with axis ratios $q_2/q_1$ and $q_3/q_2$ that fall
  in the range (0.65, 0.75) and (0.45, 0.55), respectively. All points have been colour-coded
  by the formation time variable $\log (t_{50}/t_{\rm id})$, according to the colour-bar on the 
  right. The solid line in each panel shows the collapse threshold predicted by the classic
  EC model described by BM96, which assume an initially 
  spherical perturbation geometry. The dot--dashed lines show the model predictions assuming
  initial perturbation shapes of $q_2/q_1=q_3/q_2=0.7$ (middle) and 0.5 (right) (referred to 
  as the ECE model). Horizontal dashed lines show the SC threshold, 
  $\delta_{\rm sc}=1.686$. Note that the ellipsoidal collapse threshold obtained for spherical 
  perturbations provides a poor description of the distribution of the linear overdensities of 
  non-spherical systems, particularly for the most recently collapsed ones (shown using red colours 
  on this colour-scale).}
\label{Fig:ShapeBarrier}
\end{figure*}

With estimates of the shapes, overdensities and tidal ellipticities of 
protohaloes, we can test the shape dependence of barrier heights predicted 
by the ECE model discussed above.
The distribution of protohaloes in the ($\delta_{\rm h},e$) plane is shown in 
Fig.~\ref{Fig:ShapeBarrier} for objects in three bins of initial shape. 
For simplicity, we only consider haloes with $\nu\approx 1$ but note that 
our conclusions hold for other halo masses and at all redshifts. From left 
to right, panels correspond to protohaloes with increasing triaxiality. 
The left-hand panel plots approximately spherical protohaloes, selected to have 
$q_3/q_1>0.8$. The middle and right-hand panels show triaxial systems whose 
axis ratios ($q_2/q_1$ and $q_3/q_2$) fall in the ranges 0.65 to 0.75, and 
0.45 to 0.55, respectively. As in Fig. \ref{Fig:initial_delta_e}, we have 
coloured points according to the logarithm of the half-mass formation time 
measured relative to the age of the Universe at $z_{\rm id}$: $\log \, (t_{50}/t_{\rm id})$.

As discussed in the previous section, protohalo shapes have a significant 
effect on the density contrast required for collapse. Compared to their 
spherical counterparts, non-spherical protohaloes require higher density 
contrasts to collapse when no tidal fields are present, but collapse at 
lower density contrasts in the presence of strong external tides. They 
exhibit a range of tidal ellipticies over which $\delta_{\rm ec}\approx \delta_{\rm sc}=1.686$ 
that broadens with increasing (initial) departure from sphericity. This explains 
why the triaxial protohaloes in Fig.~\ref{Fig:ShapeBarrier} tend to move to 
higher values of $e$, but not to higher $\delta_{\rm h}$, as expected from the 
standard EC model.

The solid curves in each panel of Fig.~\ref{Fig:ShapeBarrier}, for example, shows the 
ellipsoidal collapse threshold for initially spherical protohaloes (for 
$p=0$), which provides a good description of the minimum overdensities of 
spherical systems (left-most panel), particularly those with recent formation
times (red points). Note however, that this curve describes rather poorly the 
measured overdensities of the non-spherical protohaloes shown in the other two panels. 
Never the less, when the initial shapes of protohaloes are included 
in the calculation of $B_{\rm ec}$, the ellipsoidal model again provides a 
good description of their minimum Lagrangian overdensities. The dot--dashed 
curves in Fig.~\ref{Fig:ShapeBarrier} show $B_{\rm ec}$ computed for protohaloes 
with $q_2/q_1=q_3/q_2=0.7$ (middle panel) and 0.5 (right-hand panel), which trace 
rather well the minimum locus of points in each panel. Finally, note that, 
in all shape bins, the most recently collapsed haloes (shown as red points 
on this colour scale) neatly trace the ellipsoidal collapse barrier provided 
their initial shapes have been properly modelled.

\begin{figure}
\begin{center}
\resizebox{7.5cm}{!}{\includegraphics{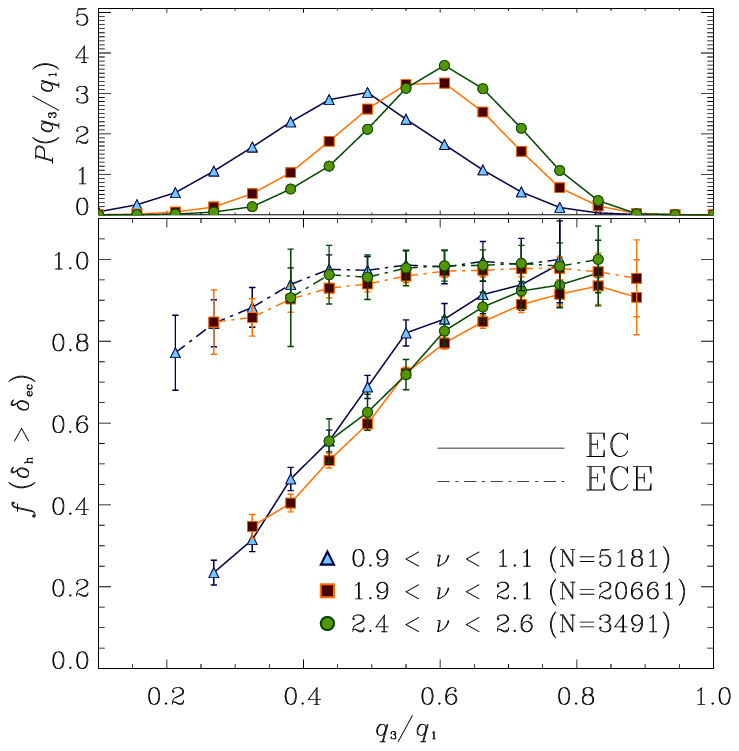}}
\end{center}
\caption{Fraction of protohaloes as a function of ``sphericity'' ($q_3/q_1$) with linear 
  density contrasts that exceed the ellipsoidal collapse threshold, $\delta_{\rm ec}$. Different 
  symbols correspond to different masses, characterized here using the peak-height mass
  parameter, $\nu$ (note that we only consider protohaloes of $z_{\rm id}=0$ systems here). 
  Different line styles are used to distinguish collapse thresholds computed for initially
  spherical perturbations (solid) from those that use the measured protohalo shapes 
  (dot--dashed lines).}
\label{Fig:frac_delta_lt_B}
\end{figure}

These results imply that the standard model for ellipsoidal collapse 
outlined by BM96 provides a poor description of the Lagrangian 
overdensities of non-spherical protohaloes. We quantify this further 
in Fig.~\ref{Fig:frac_delta_lt_B}, where we plot the fraction of 
haloes for which $\delta_{\rm h}(z_{\rm id})\geq \delta_{\rm ec}$ as a function of the 
minor-to-major axis ratio of the protohalo. For clarity, we only 
consider protohaloes of $z_{\rm id}=0$ haloes, but show results for 
objects in three separate bins of peak height: $\nu\approx 1$, 2 
and 2.5. In the upper panels we plot the distribution of $q_3/q_1$ 
for each $\nu$ bin. Note that $B_{\rm ec}$ has been computed individually 
for each halo using their measured $e$ and $p$, assuming either a spherical 
protohalo geometry (solid lines), or using the protohalo's measured initial 
shape (dot--dashed lines).

Assuming spherical symmetry when calculating $\delta_{\rm ec}$, the fraction of 
protohaloes that fall below the predicted collapse threshold depends 
sensitively on their initial shapes. Only those with initially spherical 
geometries are denser than the EC threshold given the initial strength of 
their surrounding tidal field. This is perhaps not surprising given that 
this model {\em assumes} initially spherical perturbations. Below $a_3/a_1\sim 0.7$ 
the fraction of haloes with $\delta_{\rm h}>\delta_{\rm ec}$ drops dramatically;
only $\sim 40$\% of those with $a_3/a_1\sim0.4$ satisfy this restriction. 
However, when their initial shapes have been included in the calculation of the 
collapse threshold virtually all protohaloes have $\delta_{\rm h}>B_{\rm ec}$. Protohalo 
shapes are therefore an important ingredient in analytic models that attempt to 
relate regions of the Lagrangian density field to the statistics of Eulerian 
haloes at later times. Note also that both trends are approximately independent 
of halo mass: the three bins of peak height used in Fig.~\ref{Fig:frac_delta_lt_B} 
span nearly two orders of magnitude in halo mass.

\section{Discussion}
\label{sec:discussion}

The average linear overdensity of DM protohaloes increases 
with tidal-field strength in a way that resembles the predictions
of the EC model of BM96
\citep[see also,][]{Dalal2008a,Robertson2009,Elia2012}. 
By following the collapse of individual protohaloes, and 
matching them to non-linear objects identified at later redshifts $z_{\rm id}$, we 
have shown that the scatter in $\delta_{\rm h}(z_{\rm id})$ correlates strongly with 
the halo's ``formation time'', $z_{50}$, at which half of its final 
mass was first assembled into one progenitor. Intriguingly, our results 
indicate that most recently formed haloes do not trace the classic ellipsoidal 
collapse barrier but fall systematically below it. 

By modifying the classic EC model to follow the collapse of an initially 
triaxial perturbation (rather than a spherical one) one can approximately capture 
the {\em minimum} overdensities of protohaloes as a function of $e$. However, 
in this case a conceptual problem arises from the fact that our model 
predicts rather well the minimum overdensities of collapsed regions, 
rather than the mean as is the case for the BM96 model. 
In EPS or peaks theory, one assumes that a given mass element at 
${\bmath{x}}$ will be associated with a halo of mass $M$ at redshift 
$z_{\rm id}$ provided its overdensity, smoothed with a filter of mass $M$ and 
linearly extrapolated to that redshift, has the value $\delta_M({\bmath{x}},z_{\rm id})\equiv B(z_{\rm id})$. 
Why, then, do haloes with $\delta_{\rm h}(z_{\rm id}) > B$ exist when there is 
certainly a larger mass scale $M'>M$ for which 
$\delta_{M'}({\bmath{x}},z_{\rm id})\equiv B(z_{\rm id})$ is satisfied?

One possibility is that evolutionary effects truncate mass accretion 
on to haloes at redshifts $z>z_{\rm id}$. This is possible, for 
example, if haloes were previously associated with and expelled from 
a more massive neighbouring system \citep[e.g.][]{Ludlow2009,Li2013}, or if their
accretion was halted by tides from nearby structures \citep{WangMoJing2007,Hahn2009,Behroozi2013}.
Such tidal truncation of halo growth complicates the comparison of measured 
protohalo overdensities with the model-predicted thresholds in a couple 
of ways. First, the collapse redshift of the halo will coincide with 
the time at which mass accretion halted rather than the time at which 
the halo was identified, which raises the relevant model barrier by a 
factor $\sim D(z_{\rm id})/D(z_{\rm c})$. Secondly, the tidal forces associated with 
strong interactions with neighbouring systems are not properly accounted 
for in the EC model. Such forces would inhibit collapse, further raising the 
threshold for halo formation. Incorporating such effects will
likely require a substantial reform of the ellipsoidal model.

\begin{figure*}
\begin{center}
\resizebox{12cm}{!}{\includegraphics{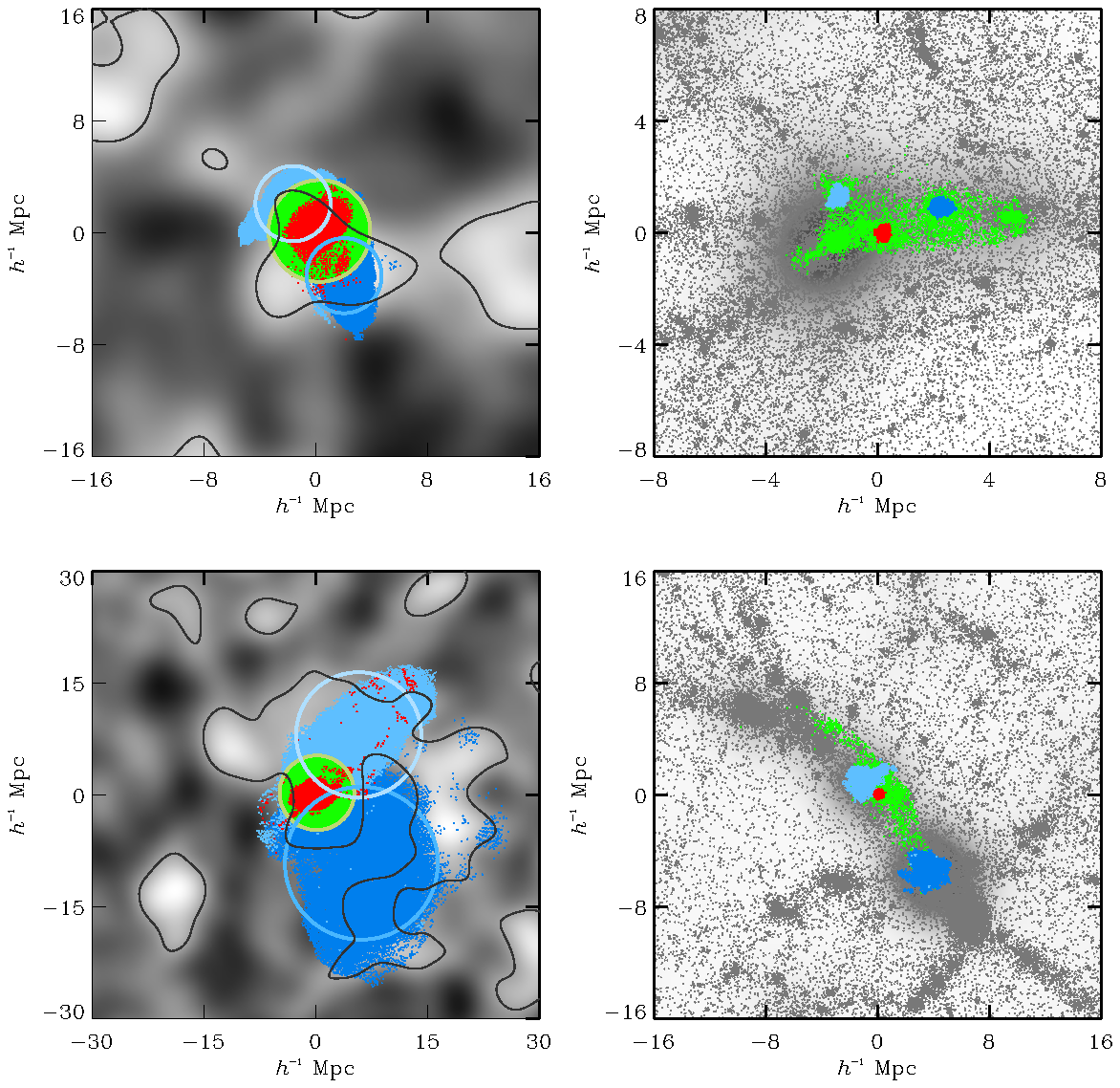}}
\end{center}
\caption{Red points show two examples of the Lagrangian (left) and 
  Eulerian (right) locations of haloes whose linear density contrasts 
  exceed the collapse threshold predicted by our ECE model for their
  measured initial shapes and external tides. Top panels correspond to
  a halo of mass $\sim$ 4.2$\times 10^{12} \, h^{-1} \, M_{\odot}$; 
  bottom panels to a $\sim$ 6.8$\times 10^{12} \, h^{-1} \, M_{\odot}$
  halo. In both left-hand panels, 
  the linear density fields have been smoothed with a tophat filter 
  scaled to the mass of the halo. Green points show the particles that 
  fill a Lagrangian sphere, centred on each protohalo, whose mean density 
  contrast is equal to the value predicted by our ECE model. Light and dark 
  blue points show neighbouring protohaloes and their descendants whose 
  growth substantially influences the collapse of the central object: these 
  haloes generate strong tidal forces that strip matter from the central halo's 
  vicinity before it is able to accrete, thereby suppressing mass accretion on to 
  the halo at early times. At $z=0$, this material is distributed amongst several 
  nearby haloes, or within filamentary structures in which the haloes form.}
\label{Fig:StrippedHaloes}
\end{figure*}

In Fig.~\ref{Fig:StrippedHaloes} we provide a couple of visual examples 
of protohaloes in our 150 \Mpch~ box that lie considerably above the 
collapse barrier inferred from our ECE model. The left-hand panels plot the linear 
density fields in the protohalo vicinity (smoothed on the mass scale of the halo) 
and panels on the right show their descendants in the evolved density fields. In 
each panel orange points show the particles that, at $z=0$, belong to the 
FoF halo. Green points are those within a Lagrangian 
sphere centred on the protohalo that encloses a mean overdensity equivalent 
to the collapse threshold predicted by our ECE model (indicated by a thick 
green circle in each panel); these highlight the mass envisioned by EPS theory 
to collapse onto the halo by $z=0$.

The first example, plotted in the top panels of Fig.~\ref{Fig:StrippedHaloes}, 
centres on a FoF halo of mass $\sim$ 4.2$\times 10^{12} \, h^{-1} \, M_{\odot}$; 
its predicted mass exceeds this value by nearly an order 
of magnitude. In the right hand panel 
it is clear that at $z=0$ much of this mass is actually distributed 
amongst several haloes that trace a large-scale filamentary structure. 
The sets of light and dark blue points in each of the top panels show the 
locations of two neighbouring haloes that have accreted $\sim 30$ per cent 
of the mass that EPS would have associated with the central object. 
The centres of mass of these neighbouring protohaloes lie outside of the 
Lagrangian sphere of overdensity $\delta_{\rm ec}$, yet their own 
spheres (shown using blue circles) overlap with it. The remaining $\sim 70$ per cent 
of the mass is today distributed within other nearby haloes, 
or diffusely within the filamentary structure.

A similar example, shown in the lower panels of Fig.~\ref{Fig:StrippedHaloes}, 
centres on a FoF halo of mass of $M \sim 6.8\times 10^{12} \, h^{-1} \, M_{\odot}$.
Based on its linear overdensity (smoothed on the scale $M$),
$\delta_{\rm h}\approx 4.2$, EPS theory and the ellipsoidal barrier would 
have associated this halo with a mass 
$M_{\rm ec}\sim 4.0\times 10^{13} \, h^{-1} \, M_{\odot}$ (green points), 
roughly a factor of 6 larger than its true mass. The lower right-hand 
panel shows that much of this mass is, at $z=0$, spread across a large, 
$\sim 10 \Mpch$ filament. This tidal stretching of the linear perturbation 
is induced by two much larger neighbouring fluctuations, one with a mass of 
$\sim 3,5\times 10^{14} \, h^{-1} \, M_{\odot}$ (light blue) and the 
other of $\sim 1.9\times 10^{12} \, h^{-1} \, M_{\odot}$ (darker blue). 
Although these are distinct structures (their locations and Lagrangian 
sizes indicated using blue circles), together they accrete $\sim 50$ per cent
of the total mass that {\em should} have collapsed upon the central halo. 

Clearly the collapse of perturbations in Gaussian random fields is 
much more complex than what is envisioned by simple dynamical models, such as
the SC or EC model. Fluctuations exist on a variety of 
spatial scales, and large-scale overdensities of different height and mass often 
overlap with one another. As a result, high density peaks capable of collapsing 
under their own self-gravity may strongly influence the evolution of neighbouring 
systems in ways that are not accounted for in the EC model. Extensions or refinements 
of the ellipsoidal model that incorporate the effects of non-linear tides, or attempt 
to model the unique assembly histories of DM haloes, may help reconcile the 
model predictions and the collapse dynamics of individual objects.

\section{Summary}
\label{sec:summary}

We have used two cosmological simulations of structure formation to study the 
properties of the Lagrangian progenitors of CDM haloes (or 
``protohaloes''), the results of which were compared directly against the 
predictions of the EC model. Our main results can be 
summarized as follows.

\begin{itemize}

\item The Lagrangian patches that collapse to form CDM haloes that are identified at redshift 
  $z_{\rm id}$ are, in general, non-spherical. Over the range of peak heights
  $0.6\simlt \nu \simlt 4$, for example, the average minor-to-major axis 
  ratio varies from roughly $\sim 0.4$ to $\sim 0.7$, and fewer than 2 per cent of 
  $\nu>1$ haloes have a ``sphericity'' $q_3/q_1>0.8$. 

\item The non-spherical shapes of protohaloes are likely determined by the 
  relative compression factors along their principal axes due the surrounding tidal field
  (see Fig.~\ref{Fig:shape_tides}). For example, haloes that form in 
  regions of the linear density field where $\lambda_1$ is large and positive, 
  but $\lambda_3$ is large and negative, accrete mass from very flattened and 
  elongated regions due to the differing gravitational tides along 
  orthogonal directions. This results in a strong alignment between the 
  principal axes of the inertia tensor of protohaloes and those of the 
  linear tidal field \citep[see also][]{Lee2000,Porciani2002b}.

\item The average overdensity of DM protohaloes, $\delta_{\rm h}$, 
  linearly extrapolated to their identification redshift, $z_{\rm id}$, 
  increases with the strength of the surrounding tidal field in a way 
  that is described reasonably well by the classic EC
  model of BM96. At fixed mass, however, the scatter in $\delta_{\rm h}$ depends strongly 
  on the formation epoch, $z_{50}$, defined as the redshift at which half of 
  its final mass was first assembled into one main progenitor. Furthermore,
  haloes that have collapsed very recently (for example, those with $t_{50}/t_{\rm id}>0.8$,
  where the $t$s are the corresponding cosmological times) tend to lie below the EC barrier 
  but are neatly bounded beneath by the SC threshold, $\delta_{\rm sc}\approx 1.686$, 
  independent of the strength of their surrounding tidal field.
  
\item This highlights a failure of the classic ellipsoidal model, which 
  systematically over-predicts the density thresholds required for 
  collapse to occur by $z_{\rm id}$. This result can be explained by
  abandoning the model's assumption 
  that Lagrangian protohaloes are spherical. Allowing for initial asymmetry
  in the perturbation shape alters the collapse times of its axes, 
  resulting in a modification to the density threshold required for 
  collapse to occur by a given time. When tuned to match the typical
  shapes of protohaloes in our simulation initial conditions, the modified 
  ellipsoidal model accurately reproduces their minimum linear over-densities.
  EC models that assume initially spherical perturbations \citep[BM96;][]{Sheth2001} 
  fail to describe the measured overdensities of protohaloes with non-spherical geometries.

\item A caveat of our model is that the majority of protohaloes have 
  $\delta_{\rm h}(z_{\rm id})\geq \delta_{\rm ec}(z_{\rm id})$, suggesting that EPS 
  theory (if executed using our collapse model) would substantially over-predict 
  the masses of most haloes. We speculate that
  this is intimately related to differences between the hierarchical growth of CDM haloes 
  and the simple collapse dynamics envisioned by ellipsoidal model. In particular, protohaloes 
  that lie considerably above our model-predicted barrier tend 
  to be highly clustered and reside in high-density regions, where overdensities
  on a variety of spatial scales likely overlap. The Lagrangian 
  state of these systems therefore differs substantially from what is 
  envisioned by simple models for the collapse of isolated density 
  perturbations. Such complications are expected 
  to affect halo collapse times, as well as the evolution of the tidal field 
  acting upon each system, and therefore complicates a direct comparison of the 
  model predictions to the outcome of numerical simulations. We will discuss
  these effects in more detail in a forthcoming paper.

\end{itemize}

Overall, our results indicate that the standard EC
model \citep[e.g. BM96;][]{Sheth2001} provides a rather incomplete census of the 
possible collapse thresholds for the formation of DM haloes from
Gaussian random fields. In addition to the specifics of the linear 
tidal field at a given location, the collapse threshold depends sensitively 
on the assumed shape of the Lagrangian region that eventually collapses 
to form a bound object, as well as on the its formation redshift. 
These results have important implications for understanding the 
origin of DM haloes; for modelling their assembly histories;
as well as for interpreting the age--dependence of halo clustering.

\section*{Acknowledgments}
ADL acknowledges financial support from the Deutsche Forschungsgemeinschaft through
the SFB (956), ``The Conditions and Impact of Star Formation'', and MB through the 
Transregio 33, ``The Dark Universe''. We thank Yehuda Hoffman and Ravi Sheth for useful
discussions, and our referee, Aseem Paranjape, for a useful report which has improved this
work.

\appendix
\section{A General Ellipsoidal Collapse Model}
\label{sec:ellipsoidal}

Following BM96, we consider a uniform density perturbation on 
top of a flat Friedmann-Robertson-Walker background whose energy content 
is dominated by the matter density, $\rho_{\rm m}$, and a cosmological 
constant, $\Lambda$. We model the perturbation as a homogeneous 
ellipsoid with semi-axes of physical length $r_i$ ($i=1,2,3$) and 
density contrast, $\delta$. We adopt a Cartesian coordinate system, 
$x_i$, that coincides with the principal axes of the ellipsoid,
which are also assumed to be perfectly aligned with the external 
tidal shear (see, e.g., Fig.~\ref{Fig:alignment}).

In this coordinate system, the internal gravitational potential of the 
ellipsoid (which satisfies Poisson's equation 
$\nabla^2 \phi=4\pi G\,\rho_{\rm m}\,(1+\delta)$) can be written
\begin{equation}
\phi_{\rm ell}({\mathbf{x}})=\pi \, G \,(1+\delta)\,\rho_{\rm m}\,
\sum_{i=1}^3 b_i x_i^2\;,
\label{eq:carlson}
\end{equation}
where the coefficients $b_i$ are given by Carlson's elliptic integrals \citep{Kellogg1929,Chandrasekhar1969}:
\begin{equation}
b_i=r_1\, r_2 \,r_3 \int_0^\infty \frac{d\tau}{(\tau+r_i^2)
\displaystyle{\prod_{j=1}^3}(\tau+r_j^2)^{1/2}}.
\label{ellpot}
\end{equation}
The potential generated by the uniform distribution of matter 
outside the perturbation is given by
\begin{equation}
\phi({\mathbf x})=\pi \, G \, \rho_{\rm m} \, \sum_{i=1}^3\left(\frac{2}{3}-b_i\right)\,x_i^2 -\frac{\Lambda c^2}{6} \,\sum_{i=1}^3 x_i^2.
\label{eq:phiouter}
\end{equation}
The first term in this expression is the external potential generated 
by the perturbation itself. It is obtained by considering the superposition 
of a uniform matter distribution and an ellipsoidal ``hole'' with a density 
of $-\rho_{\rm m}$. The second term in equation~(\ref{eq:phiouter}) is the contribution 
from the cosmological constant. The total gravitational potential is thus given by
\begin{equation}
\phi({\mathbf x})=\pi \, G \, \rho_{\rm m} \, \sum_{i=1}^3\left( \left[ (1+\delta)\,b_i-\beta_i \right]  -\frac{\Lambda c^2}{6} \right) \,x_i^2\;,
\end{equation}
where we have used $\beta_i=b_i-2/3$.

Assuming that the tidal shear is parallel to the principal axes of the 
density perturbation (which naturally arises when one starts with a sheared 
spherical perturbation), we can write the equation of motion for the 
semi-axis lengths $r_i$ of the ellipsoid as
\begin{equation}
\frac{\ddot{r}_i}{r_i}=\frac{\Lambda c^2}{3}-4\,\pi \, G \,\rho_{\rm m} \left(\frac{1+\delta}{3}
+\frac{\beta_i}{2}\,\delta+\lambda^{\rm ext}_i \right)\;,
\label{eqbm96} 
\end{equation}
where dots denote time derivatives and we have used the fact that
$b_i\,(1+\delta)-\beta_i=\beta_i\,\delta+2\,(1+\delta)/3$.

Because the potential is quadratic and the acceleration is linear in the 
coordinates, concentric ellipsoids with the same axis ratios evolve 
self-similarly and the density perturbation remains homogeneous at all 
times. Given the comoving coordinates of the axes extrema, $q_i$, the 
evolution thus preserves $r_1\,r_2\,r_3\,(1+\delta)=q_1\,q_2\,q_3\,a^3=\,$constant, 
and the overdensity of the perturbation evolves as
\begin{equation}
\delta(a)=\frac{q_1\,q_2\,q_3}{r_1\,r_2\,r_3}\,a^3-1\;.
\end{equation}

Initial conditions for equation~(\ref{eqbm96}) can be set according to the Zel'dovich 
approximation at some early time, $t_0$:
\begin{eqnarray}
r_i(t_0)&=&q_i \,\left[1-\lambda_i(t_0)\right]\, a(t_0)\;,
\label{zeld}\\
\dot{r}_i(t_0)&=&H(t_0)\,r_i(t_0)-q_i\,H_D(t_0)\,
\lambda_i(t_0)\, a(t_0)\;.
\end{eqnarray}
Here $H=\dot{a}/a$ is the Hubble parameter; $D$ the linear growth factor; 
$H_D\equiv\dot{D}/D$, and $\lambda_i(t)\propto D(t)$ are the eigenvalues 
of the linear deformation tensor. In order to match the last expressions 
with equation~(\ref{eqbm96}), we also derive the eigenvalues of the deformation 
tensor at $t_0$ from the peculiar gravitational potential (which satisfies 
$\nabla^2 \Phi =\delta$):
\begin{eqnarray}
\lambda_i^{\rm tot}(t_0)&=& \frac{\delta(t_0)}{2}\,b_i(t_0)+\lambda_i^{\rm ext}(t_0) \\ \nonumber
&=&\delta(t_0) \left(\frac{1}{3}+
\frac{\beta_i(t_0)}{2}\right)+\lambda_i^{\rm ext}(t_0)\;.
\label{match}
\end{eqnarray}
Only the components of this expression that scale as $D(t)$ will contribute 
to the $\lambda_i$ terms appearing in the Zel'dovich approximation.

The early evolution of the internal gravitational potential of the ellipsoid 
can be determined by combining equation~(\ref{ellpot}) and (\ref{zeld}) and 
Taylor expanding in powers of $D(t)$:
\begin{equation}
\beta_i(t)=
\beta_i^{(0)}+\Delta \beta_i\,D(t)+{\cal O}[D^2(t)]\;.
\end{equation}
Here $\beta_i^{(0)}$ is evaluated using equation~(\ref{ellpot}) with $r_i=q_i$,
and the second term in is given by
\begin{equation}
\Delta \beta_i\,D(t)=\sum_{j=1}^3\alpha_{ij}\,\lambda_j,
\end{equation}
where
\begin{equation}
\alpha_{jj}= q_1 \,q_2 \,q_3 \int_0^\infty
\frac{(2 q_j^2 - \tau)\,d\tau}{(q_j^2 +
     \tau)^2 \displaystyle{\prod_{i=1}^3 (q_i^2 + \tau)^{1/2}}}\;
\end{equation}
and, for $j\neq i$, 
\begin{equation}
\alpha_{ji}= -q_1 \,q_2 \,q_3 \int_0^\infty
\frac{\tau \,d\tau}
{(q_j^2 + \tau) (q_i^2 + \tau)
 \displaystyle{\prod_{i=1}^3 (q_i^2 + \tau)^{1/2}}},
\end{equation}
When all $q_i$s are equal $\alpha_{ij}=8/15$ for $i=j$ and 
$-4/15$ when $i\neq j$ and this reduces to the result of BM96, 
$\beta_i\simeq\Delta \beta_i\,D(t)=(4/5)\,t_i$. Thus, the term 
$\beta\, \delta/2$ in equation~(\ref{match}) is of second order in $D(t)$ and does not 
contribute to the linear shear tensor, which is fully generated by the external
tides. For initially triaxial perturbations, however, the linear velocity shear 
is composed of two terms: 
\begin{equation}
\lambda_i=\frac{\delta}{3}+\biggr[\frac{\beta^{(0)}_i\,\delta}{2}+\lambda_i^{\rm ext} \biggl].
\end{equation}
For the typical axes ratios of protohaloes in our simulations 
($q_2/q_1\simeq q_3/q_2\simeq 0.8$) we find $\beta^{(0)}_i\simeq(-0.170, -0.011, 0.182)$,
showing that internal shear slows down collapse along the long axis and speeds 
it up along the short axis.

Solving equation~(\ref{eqbm96}) requires knowledge of the time evolution of the 
external tidal field, $\lambda^{\rm ext}_i$. BM96 proposed two 
limiting approximations. In one case, the external tides are assumed to 
grow independently of the ellipsoid and the shear scales as the linear 
growth factor,
\begin{equation}
\lambda^{\rm ext}_i(t)=\frac{D(t)}{D(t_0)}\, \lambda^{\rm ext}_i(t_0)\;.
\label{lintide}
\end{equation}
Alternatively, one can assume that the external shear is dominated by the 
shape of the ellipsoid itself. In this case, $\lambda^{\rm ext}_i(t)=(5/4)\, \beta_i$ 
(note that this reduces to equation~(\ref{lintide}) at early times), or, more generally 
for triaxial perturbations,
\begin{equation}
\lambda^{\rm ext}_i(t)=\sum_j \alpha^{-1}_{ij}
\left(\beta_j-\beta_j^{(0)} \right)\;,
\label{nonlintide}
\end{equation}
where the $\alpha^{-1}_{ij}$ are the elements of the inverse matrix of 
$\alpha_{ij}$. A hybrid model which interpolates between these asymptotic 
regimes has been recently proposed by \citet{AngrickBartelmann2010}. Here
one uses the non-linear growth proposed by BM96 until the 
corresponding axis turns around and linear growth is used afterward. 

Finally, the equations of motion for the axis lengths do not account for the physics 
of virialization. Relaxation processes are instead mimicked by ``freezing'' collapse 
along each axis once a critical radius $r_{{\rm eq},i}=f\,a\,q_i$ is reached. 
BM96 suggested using the radial freeze-out factor $f=0.178$, which results
in a virial density contrast of $\sim 178$ in the SC model;
this value is also commonly adopted in applications of the EC model.
More recently, \citet{AngrickBartelmann2010} used the tensor 
virial theorem to predict more realistic values of $f$. However, the last phase of 
collapse is generally rapid, and altering the value of $f$ does not strongly affect the 
estimated epoch of virialization. For ease of comparison with previous work, 
all applications of this model presented in this paper have assumed a freeze-out factor 
$f=0.178$.

\bibliographystyle{mn2e}
\bibliography{paper}

\end{document}